\documentclass[letter]{aa} 
\usepackage{graphicx}
\usepackage{txfonts}
\usepackage{natbib}
\bibpunct{(}{)}{;}{a}{}{,}

\begin{document}

\title{Evidence for the disintegration of KIC 12557548 b}

\author{M.~Brogi\inst{1} \and C.~U.~Keller\inst{1} \and M.~de Juan Ovelar\inst{1} \and M.~A.~Kenworthy\inst{1} \and R.~J.~de Kok\inst{2} \and M.~Min\inst{3} \and I.~A.~G.~Snellen\inst{1}}

\institute{
  Leiden Observatory, Leiden University,
  P.O. Box 9513, 2300RA Leiden, The Netherlands\\
  \email{brogi@strw.leidenuniv.nl}\label{inst1}
  \and 
  SRON, Sorbonnelaan 2, 3584 CA Utrecht, The Netherlands\label{inst4}
  \and
  Astronomical Institute Anton Pannekoek, University of Amsterdam,
  Postbus 94249, 1090GE Amsterdam, The Netherlands\label{inst3}
}

\date{Received / Accepted }

\abstract
   {The $Kepler$ object KIC 12557548 b is peculiar. It exhibits transit-like features every 15.7 hours that vary in depth between 0.2\% and 1.2\%. Rappaport et al. (2012) explain the observations in terms of a disintegrating, rocky planet that has a trailing cloud of dust created and constantly replenished by thermal surface erosion. The variability of the transit depth is then a consequence of changes in the cloud optical depth.}
   {We aim to validate the disintegrating-planet scenario by modeling the detailed shape of the observed light curve, and thereby constrain the cloud particle properties to better understand the nature of this intriguing object.}
   {We analysed the six publicly-available quarters of raw $Kepler$ data, phase-folded the light curve and fitted it to a model for the trailing dust cloud. Constraints on the particle properties were investigated with a light-scattering code.}
   {The light curve exhibits clear signatures of light scattering and absorption by dust, including a brightening in flux just before ingress correlated with the transit depth and explained by forward scattering, and an asymmetry in the transit light curve shape, which is easily reproduced by an exponentially decaying distribution of optically thin dust, with a typical grain size of 0.1~$\mu$m.}
   {\rm Our quantitative analysis supports the hypothesis that the transit signal of KIC 12557548 b is due to a variable cloud of dust, most likely originating from a disintegrating object.}


\maketitle

\section{Introduction}

\citet{rap12} recently reported the discovery of a peculiar signal in the light curve of KIC 12557548\footnote{Henceforth denoted with KIC1255 b}. A transit-like signal is seen every 15.7 hours, with a depth greatly varying from event to event, ranging from less than 0.2\% to more than 1.2\%. The transit timing is consistent with a single period and shows no variations down to the 10$^{-5}$ level. This suggests that the changes in transit depth cannot be produced by a fast precession of the orbit, which would in turn require the gravitational perturbation of a second companion in the system. In addition, the phase-folded light curve shows no evidence for ellipsoidal light variations down to 5 parts in 10$^5$. This sets an upper limit of about 3~$M_{Jup}$ on the mass of the orbiting object. The explanation proposed by \citet{rap12} entails the disintegration of a super-Mercury, caused by the high surface temperatures. Material stripped away from the planet surface forms and stochastically replenishes a cloud of dust that obscures part of the stellar disk, causing the dimming of the stellar light to randomly vary from orbit to orbit. Only a planet with mass on the order of Mercury would allow dust particles to accelerate to the escape velocity and form a trailing cloud of sufficient size.

To test the interpretation of \citet{rap12}, a quantitative understanding of the observed transit light curve is required, in particular the asymmetry of the in-transit portion of the light curve and the excess of flux before ingress. We reduced the complete set of publicly available {\it Kepler} observations, phase-folded the light curve and grouped the transits based on their depth, as explained in Sect.~\ref{data}. A model for the trailing dust, described in Sect.~\ref{model}, is then fitted to the data to estimate the transit parameters, the basic shape of the cloud and the average size of the dust particles. The results of our analysis and their interpretation are discussed in Sect.~\ref{results}.

\section{Analysis of {\it Kepler} data}\label{data}

The star KIC 12557548 was observed by {\it Kepler} in {\it long--cadence} mode, delivering one frame every 29.4 minutes. This results in only 3--4 in-transit points per orbit, and therefore only an averaged and smoothed version of the light-curve can be recovered by phase-folding the photometric time series. The public release of January 2012 extended the available data span to $\sim$1.4 years of observations (six quarters) and includes more than 22,000 photometric points. To combine and normalize the entire series, we designed a dedicated data reduction approach that starts with the raw {\it Kepler} photometry, which is obtained with simple aperture photometry. The {\it Kepler} pipeline also produces decorrelated photometry, using an automated decorrelation procedure. This, however, often delivers over-corrected light curves, in which the astrophysical signal is reduced or even cancelled together with the instrumental effects\footnote{http://keplergo.arc.nasa.gov/ContributedSoftwareKepcotrend.shtml}. Therefore the decorrelated photometry data are not suitable for our purposes.

A visual inspection of the raw light curve identifies the following features:
\begin{enumerate}
\item Jumps in flux from quarter to quarter, at the level of several percent;
\item Jumps or fast variations in flux within a quarter, at the percent level or less;
\item A periodic modulation of the stellar flux, probably due to spots on the rotating surface or other intrinsic stellar variability;
\item Transits-like features with a well-defined cadence but variable depth.
\end{enumerate}

The instrumental systematics can be efficiently modelled with a linear combination of the {\it Cotrending Basis Vectors} (CBVs), which track the state parameters of the spacecraft and the instrumentation in time\footnote{see http://archive.stsci.edu/kepler/release\_notes/release\_notes12/\\
  DataRelease\_12\_2011113017.pdf}.  
We determined the optimum number of CBVs for each quarter by minimizing the variance of the out-of-transit data, while visually checking that the periodic modulation attributed to stellar activity is preserved as much as possible. For each set of state vectors, we determined their coefficients via linear regression and divided the data by the linear model. In this way different quarters of data were combined and instrumental systematics were minimized while preserving the astrophysical signal. To remove the low-order stellar modulation, we discarded the data points belonging to the transits and averaged the remaining points in each 15.7-hour orbit. This operation was performed separately for each section of contiguous data (i.e.\ data without temporal gaps) to avoid flux discontinuities. We then spline-interpolated the averaged points in time and divided the light curve by this baseline stellar signal.

\subsection{Transit light curve}\label{folding}

\begin{figure}
\resizebox{\hsize}{!}{\includegraphics{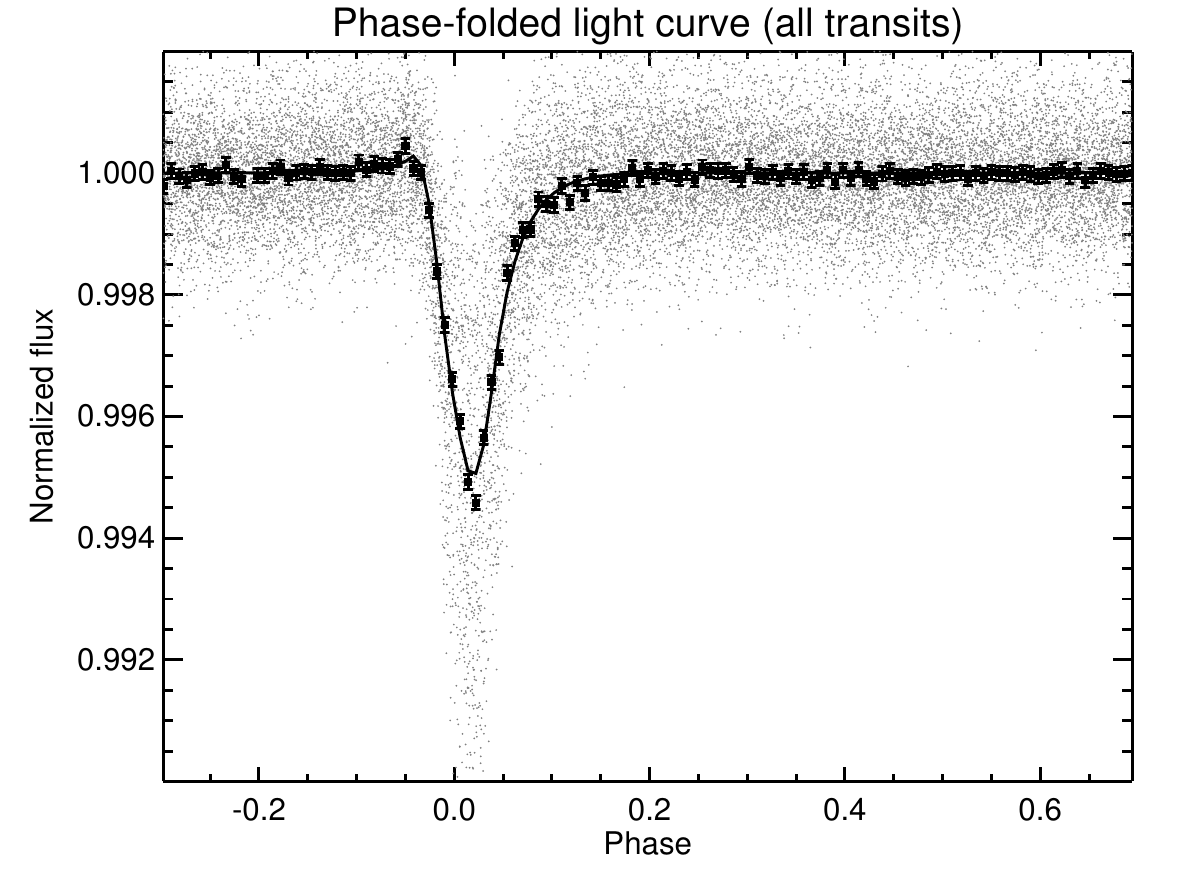}}
   \caption{The normalized light curve of KIC 1255b, phase-folded using a period of $P = 15.6854$ hours. A binned light curve (solid squares, with error bars) and the best-fit model (solid line) are overplotted.}\label{FigCurve}
\end{figure}

\begin{figure}[ht]
   \resizebox{\hsize}{!}{\includegraphics{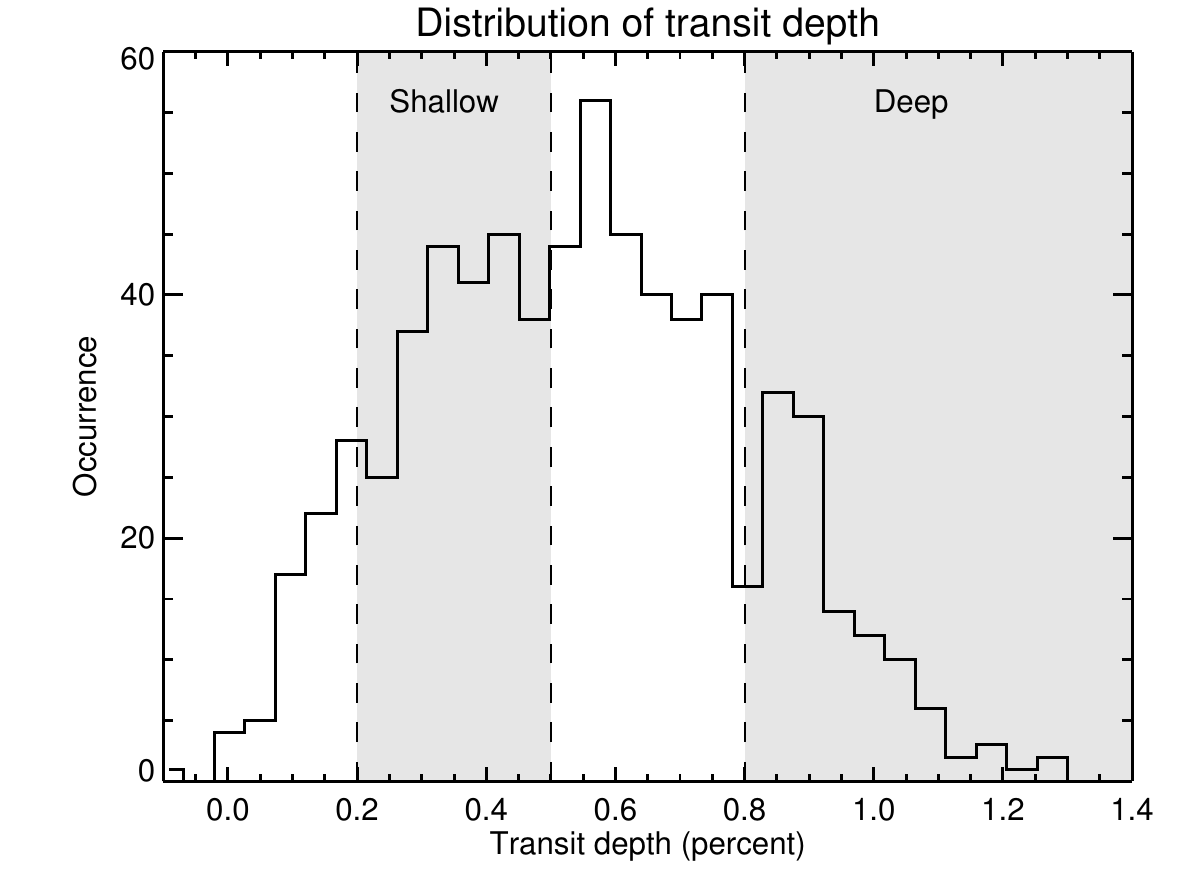}}
   \caption{Distribution of the transit depths, as measured by fitting the {\it Kepler} photometry with our model light curve with all parameters but the transit depth fixed to the best-fit values. Orbits that don't sample the in-transit portion of the curve are discarded.}\label{hist}
\end{figure}

\begin{figure*}
\includegraphics[width=8.5cm]{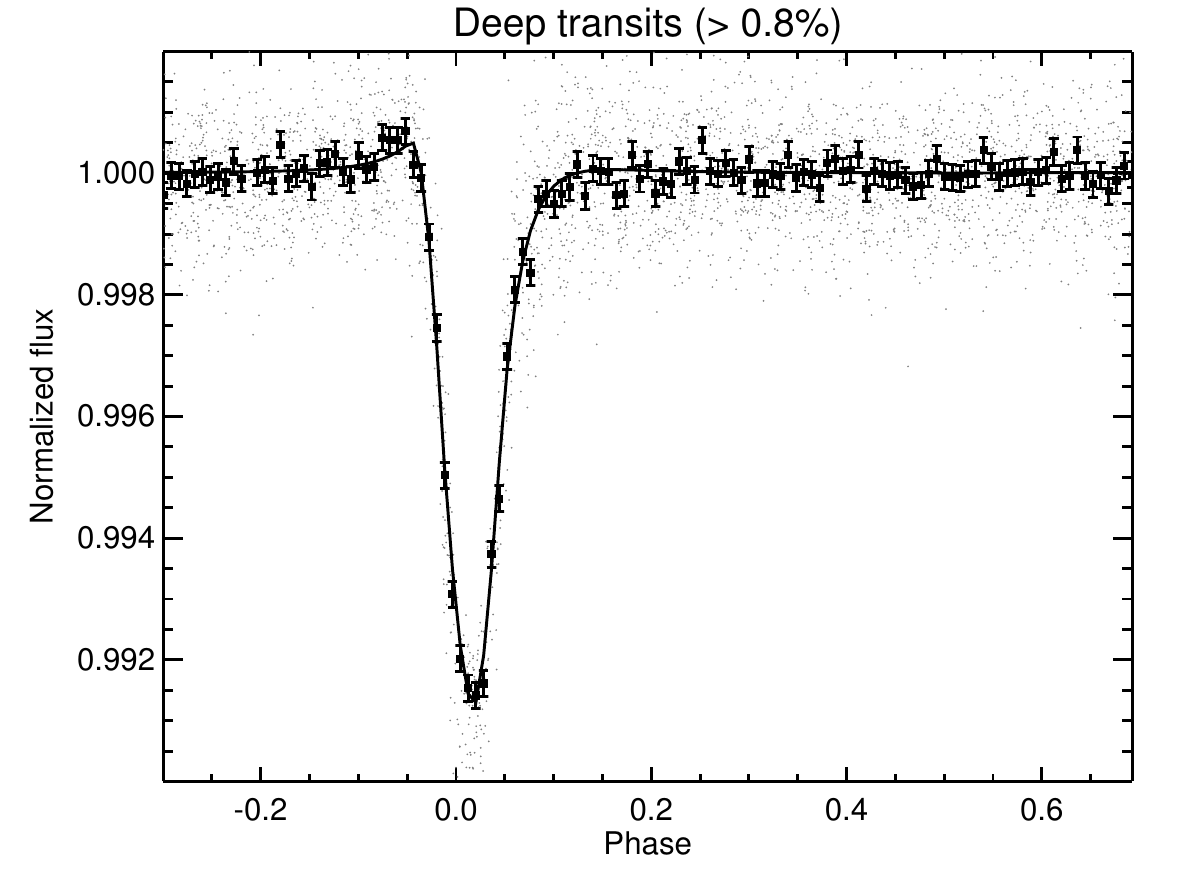}
\includegraphics[width=8.5cm]{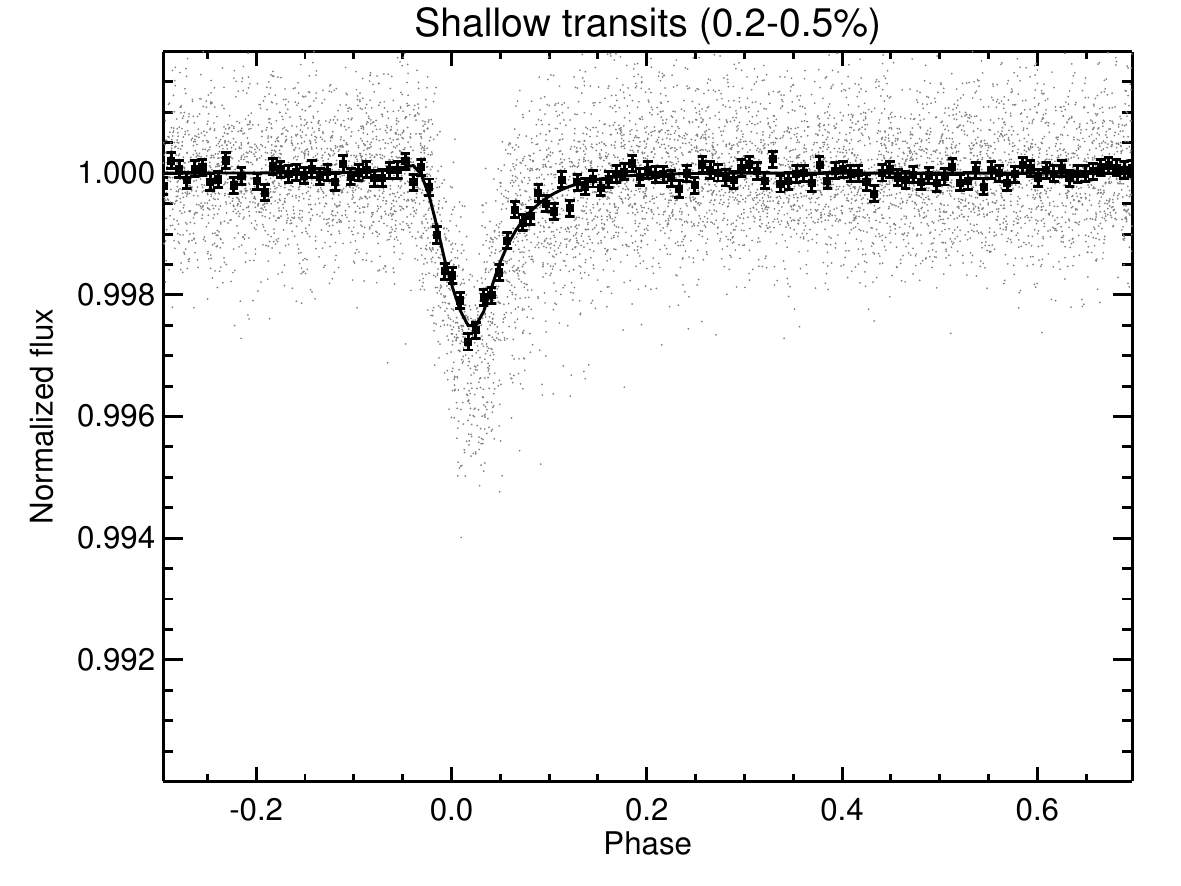}
\caption{Light curves of KIC 1255b, obtained by dividing the events into two groups based on their transit depths and binning the photometry by 0.008 in phase. The left panel shows the average light curve of the deep events (deeper than 0.8\%), and the right panel shows the average light curve of the shallow events (depths between 0.2\% and 0.5\%). The best-fit models for the two light curves are shown as a solid line.}\label{fig_bin}
\end{figure*}

We phase-folded the normalized light curve using a period of $P = 15.6854$ hours, as determined by \citet{rap12}. 
The resulting transit light curve is shown in Fig.~\ref{FigCurve} with the individual data points in light grey. 
A relative photometric precision of 8.2$\times$10$^{-4}$ per-point is achieved, which is $\sim$30\% higher than expected from pure photon noise. When binned by 0.008 in phase (black squares in Fig.~\ref{FigCurve}), 
the scatter in the data is reduced by a factor of $\sim$13, i.e.\ the square root of the number of points per bin, suggesting that systematic noise is not significantly present in the data at this level of photometric accuracy. The binned light curve exhibits the following characteristics:
\begin{enumerate}
\item An asymmetry between ingress and egress, with the latter being significantly longer than the former;
\item Absence of a secondary eclipse within the photometric precision;
\item A small increase in flux just before ingress.
\end{enumerate}
To assess the dependence of the shape of the transit light curve on transit depth, we divided the events into two extreme classes: {\it shallow} (depth between 0.2\% and 0.5\%) and {\it deep} (depth $>$~0.8\%). These ranges were chosen to maximize the difference in depth while maintaining adequate photometric precision. 

We determined the transit depths by first fitting the binned light curve, as explained in Sect.~\ref{model}, to determine the overall parameters of the transits. Subsequently, each separate orbit was fitted with the same curve, keeping all parameters fixed except for the depth, which we determined by estimating a scaling factor via a $\chi^2$ minimization. The resulting distribution of transit depths is shown in Fig.~\ref{hist}. It peaks around 0.6\% and extends from zero depth to about 1.3\%. The negative transit depths in the histogram are due to noise in the photometry. The absence of a clustering around zero depth suggests that almost every orbit observed by {\it Kepler} shows a transit-like feature, except for $\sim$8 days during quarter 2, in which the signature seems to be absent. This results in an upper limit of the planet radius, as determined in Sect.~\ref{results}.

The light curves for the deep and shallow transits, shown in Fig.~\ref{fig_bin}, share the same features of the global light curve except for the flux excess before ingress, which is not evident in the curve for the shallow events. The deep and shallow light curves, binned by 0.008 in phase, have a relative photometric precision per-point of 1.7$\times$10$^{-4}$ and 1.2$\times$10$^{-4}$ respectively. The solid lines in Fig.~\ref{fig_bin} show the best-fit model light curves, which were obtained by separately fitting the free model parameters.
\begin{figure}
\includegraphics[width=8.5cm]{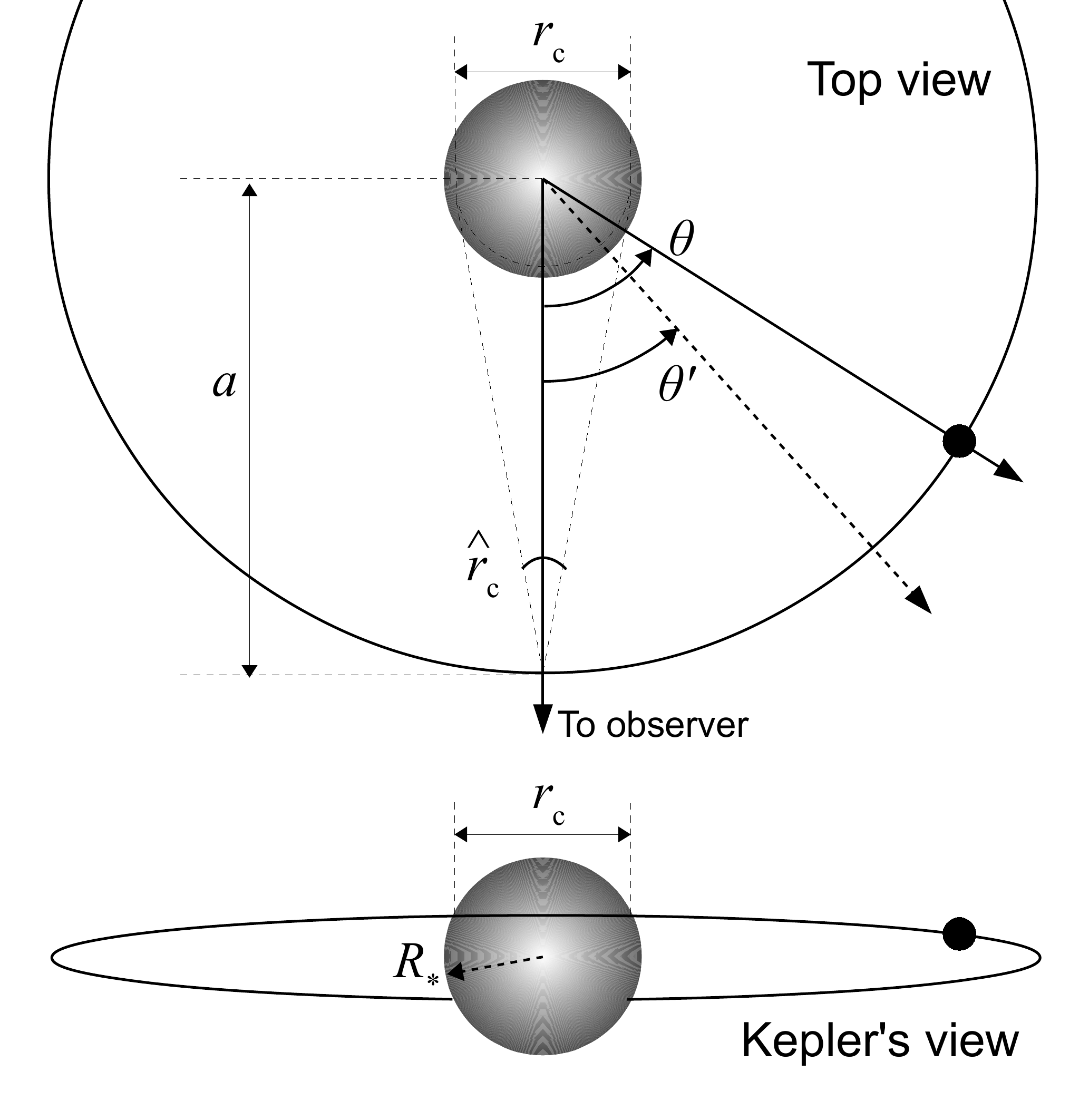}
\caption{Graphical representation of some of the quantities used in our model. The stellar and orbital radius are to scale, while the position of the orbiting body is shown with a black dot.}\label{model}
\end{figure}

\section{The dust model}\label{model}

We created a simple dust model with a minimum number of free parameters to quantitatively model the transit-like light curve. The most relevant variables are shown in Fig.~\ref{model}. We assumed a cloud of dust distributed on a circular orbit at the same radial distance as the parent body ($a$ = 0.013 AU). The latter is supposed to not contribute to the eclipse in a measurable way, as suggested by the absence of any transit-like signals during 8 days of {\it Kepler} photometry. It is therefore also reasonable to assume that the vertical extent of the cloud is negligible compared to the stellar diameter. Hence our model is one-dimensional, and it is expressed in terms of $\theta$, the angle between the observer, the center of the star and the orbiting body. This translates to an orbital phase $\varphi$ according to $\theta = 2\pi\varphi$.
The planet crosses a chord of length $r_\mathrm{c}$ on the stellar disk, corresponding to an impact parameter of $b = [1-(r_\mathrm{c}/2R_*)^2]^{1/2}$ for a stellar radius $R_*$. The crossed chord, as seen from the cloud orbit, has an angular extension of $\hat{r}_\mathrm{c}$~=~$\arcsin(r_\mathrm{c}/2a)$, which is a free parameter of the model.  

The cloud of dust is assumed to be optically thin, and its extinction cross-section $\rho$ is expressed in units of stellar area. It drops exponentially away from the planet and is parametrized by a multiplicative factor $c_\mathrm{e}$ and by an exponential parameter $\lambda$, according to
\begin{equation}
\rho(\Delta\theta)= \frac{\rho_0}{\pi R_*^2}\mathrm{e}^{-\lambda(\Delta\theta)} \equiv c_\mathrm{e} \mathrm{e}^{-\lambda(\Delta\theta)}\;,
\end{equation}
where $\Delta\theta = (\theta-\theta^\prime)$ is the angular distance between the position of the planet and an arbitrary point along the orbit, measured clockwise. The total model light-curve in units of stellar flux as a function of orbital phase $\varphi$ is
\begin{equation}
I(\varphi) = \frac{1}{\delta}\int_{2\pi\varphi-\delta/2}^{2\pi\varphi+\delta/2}{\left[1-I_\mathrm{e}(\theta)+I_\mathrm{s}(\theta)\right] d\theta},
\end{equation}
i.e. the stellar flux corrected for the extinction component $I_\mathrm{e}$ and the scattering component $I_\mathrm{s}$, convolved with a boxcar function that simulates the 29.4-minute exposure time of {\it Kepler}, which is long compared to the intrinsic temporal variations. The exposure time $\Delta t$ translates into an angular range of $\delta = 2\pi\Delta t/P$, where $P$ is the orbital period of the planet. 

The extinction component $I_\mathrm{e}(\theta)$ is given by 

\begin{equation}
I_\mathrm{e}(\theta)=\int_{0}^{2\pi}{\rho(\theta-\theta^\prime)\ i(\theta^\prime,\hat{r}_\mathrm{c})\ d\theta^\prime},
\end{equation}
which is the convolution between the extinction cross-section and the intensity of the crossed stellar disk $i(\theta^\prime,\hat{r}_\mathrm{c})$. The latter is described by the limb-darkening law $i(\mu)=1-u(1-\mu)$, where $\mu$ is the cosine of the angle between the line of sight and the normal to the stellar surface, and $u=0.79$ is the linear limb-darkening coefficient appropriate for a K4V star in the $V$-band \citep{claret00}. Expressing $\mu$ in terms of $\theta^\prime$ and $\hat{r}_\mathrm{c}$ gives
\begin{equation}
i(\theta^\prime,\hat{r}_\mathrm{c}) = 1 - u\left[1-\frac{a}{R_*}\sqrt{\sin^2(\hat{r}_\mathrm{c}/2)-\sin^2\theta^\prime}\right],
\end{equation}
set to zero for $\hat{r}_\mathrm{c}/2 < \theta^\prime < 2\pi-\hat{r}_\mathrm{c}/2$.
\\
The scattering component $I_\mathrm{s}(\theta)$ is given by
\begin{equation}\label{scatt}
I_\mathrm{s}(\theta) = \pi\varpi\left(\frac{R_*}{a}\right)^2\int_0^{2\pi}{\rho(\theta-\theta^\prime)\ \bar{p}(\theta^\prime)\ d\theta^\prime} \;,
\end{equation}
where $\bar{p}(\theta^\prime)$ is the Henyey-Greenstein (H-G) phase function \citep{henyey41}
\begin{equation}
\bar{p}(\theta^\prime)=\frac{1-g^2}{4\pi\left(1-2g\cos\theta^\prime+g^2\right)^\frac{3}{2}}\;, 
\end{equation}
set to zero for $|\pi-\theta^\prime|\le \hat{r}_\mathrm{c}/2$ to account for the secondary eclipse, and containing a single free parameter, the asymmetry parameter $g$ with $-1 < g < 1$. 

In Eq.~(\ref{scatt}) we assume that the star is a point source. Since the scattering is beamed with a cone angle $\phi$ for an asymmetry parameter $g$, and $g \simeq \cos\phi$ \citep{lam97}, the effects of the stellar disk become relevant only when $\phi\la 27^\circ$, the stellar diameter seen from the orbit of the cloud. This corresponds to $g\ga 0.89$ and therefore our best-fitting values (see Tab.~\ref{recap}) should not be significantly affected by the assumption that the star is a point source. Therefore, we do not include the integration over the stellar disk, which significantly speeds up the model computation. Finally, the single-scattering albedo $\varpi$, i.e.\ the ratio between the extinction cross-section and the scattering cross-section, is a free parameter.

The model is fitted to the binned light curves with a Markov-Chain Monte Carlo (MCMC) with a Metropolis-Hastings algorithm \citep{has70}. Sequences of $5\times10^5$ steps are generated starting from different initial conditions, and merged after discarding the first 10\% of data to account for the {\it burn-in}. The mixing between the chains is tested \citep{gel-rub92} and the $\pm$1$\sigma$ errors of the parameters are computed by measuring the interval between 15.8\% and 84.2\% of the merged distribution. The best-fit values and their uncertainties are listed in Tab.~\ref{recap}, for the average light curve and for the deep and shallow transits, respectively.

\section{Results and discussion}\label{results}

The model parameters describing the fit to the average light curve do not exhibit typical signs of bias due to the fact that we are averaging events of all depths: i) different MCMC chains mix well, indicating that the solution we find is independent of the initial parameter estimates, and ii) the model parameters are intermediate between those of the shallow and deep transit light curves, as expected if the global curve is a combination of the two. Therefore, we will start with discussing the model parameters obtained for the average light curve.


The transits of KIC1255b occur at a moderate impact parameter of $b = 0.63$. The asymmetric shape of the curve, with a sharp ingress and a slow recovery of the flux in egress, is well explained by a trailing cloud with a density that decreases away from the parent body. The characteristic length of the tail, $1/\lambda$, is about 2.5 times the diameter of the stellar disk as seen from the cloud orbit.  

The H-G asymmetry parameter $g$ being close to +1 indicates particles that strongly favor forward scattering over backward scattering.  This reproduces the observed flux excess just before ingress. The fact that this happens slightly earlier than the ingress is due to the slow {\it Kepler} cadence and the limb-darkened stellar profile. Without implementing these two in the model, it is impossible to reproduce the position of the scattering peak. Note that a similar forward-scattering contribution is present in egress, but it is masked by the dominating absorption of the tail of dust. Finally, the absence of a secondary eclipse is explained by strongly reduced backward scattering.
\begin{table}
	\caption{Best-fit parameters and their 1$\sigma$ uncertainties, as derived from the MCMC analysis. From top to bottom: impact parameter, time of mid-transit (in BJD-2,455,000), decay factor, total extinction cross-section (in units of stellar area), asymmetry parameter and single-scattering albedo.}
	\label{recap} 
  	\centering
	\begin{tabular}{|c|ccc|}
	\hline
    	Parameter & Average & Deep & Shallow \\
    	\hline
    	$b$ & 0.63$\pm0.03$ & 0.46$^{+0.02}_{-0.04}$ & 0.61$\pm0.04$ \\
    	$T_0$ & 451.9449(3)  & 451.9461(5) & 451.9430(7) \\
    	$\lambda$ & 5.1$\pm0.2$ & 6.4$\pm0.3$ & 3.86$\pm0.25$\\
    	$c_\mathrm{e}$ & 0.030$\pm0.005$ & 0.036$^{+0.003}_{-0.002}$ & 0.012$\pm0.001$ \\
    	$g$ & 0.874$^{+0.011}_{-0.019}$ & 0.743$^{+0.044}_{-0.064}$ & 0.87 (fixed) \\
    	$\varpi$ & 0.65$^{+0.09}_{-0.10}$ & $>$0.94 & 0.65 (fixed) \\
    	\hline
  	\end{tabular} 
\end{table}
\subsection{Dependence on depth}

A comparison between shallow and deep transits shows that the total extinction cross-section decreases with transit depth, and that its exponential drop is slower for shallower transits. The photometric precision of the shallow light curve is insufficient to detect the forward-scattering peak, therefore we fixed the asymmetry parameter and the albedo to $g=0.87$ and $\varpi=0.65$, respectively, as resulting from the fit of the global light curve, which has the best photometric precision. Although $\varpi$ and $g$ for the deep light curve significantly differ from those for the average light curve, we cannot determine whether or not this reflects different physical properties of the particles. In fact, in our MCMC simulations the extinction cross-section and the asymmetry parameter are moderately correlated, due to the fact that the measured transit depth is the summed contribution of the extinction and scattering components. It is impossible to solve this degeneracy with observations at a single wavelength.

\subsection{Constraints on particle size}

To constrain the particle size in the cloud, we compared realistic phase functions of silicates with the best-fitting
H-G function obtained from the MCMC analysis of the average light curve. Following \citet{boh83}, we calculated the optical properties of particles given their material \citep{dra84}, shape and size distribution. We assumed
spherical grains (i.e\ Mie scattering is the dominant process), as expected if the grains liquify before evaporating
due to the high temperature. The particle size distribution is described by a power law with a -3.5 exponent, with upper and lower limits left as free parameters.
The resulting phase functions were compared to a range of H-G phase functions constructed by varying $\varpi$ and $g$ within their $1\sigma$ uncertainties.
A distribution of particles with sizes between 0.04 and 0.19 $\mu$m best agrees with the explored range of H-G functions. Although our results are far from conclusive, the analysis suggests that the grains composing the cloud have an average size of $\sim0.1\ \mu$m.

\subsection{The size of the parent body}

As explained in Sect.~\ref{folding}, during 8 days of {\it Kepler} observations no transit-like feature is visible. The phase-folded light curve, constructed using the points in that time interval, has an in-transit photometric accuracy of 2.8~$\times$10$^{-4}$ per point. This translates into a 1$\sigma$ upper limit of 1.15~$R_\oplus$ for the radius of the parent body.

\section{Conclusions}\label{conclusion}

A simple, one-dimensional model of a trailing dust cloud can reproduce the observed light curve of KIC1255 b in exquisite detail. Features such as the flux excess before ingress, the sharp drop in ingress, the pointed shape of the flux minimum and the long flux recovery after egress are all signs of a trailing cloud of dust that periodically occults the star. The most likely origin of this trailing dust is a planetary or sub-planetary body that evaporates due to the intense stellar irradiation. If this is indeed the case, we are presented with the unique opportunity to probe the interior of an exoplanet. Although the geometry of the transit is well constrained by the currently available {\it Kepler} data, further observations with a faster cadence and multi-wavelength coverage are required to not only constrain the size but also the composition of the grains. 

\begin{acknowledgements}
We thank Eugene Chiang for a stimulating discussion about the nature and properties of KIC1255b.
\end{acknowledgements}

\bibliographystyle{aa}
\bibliography{biblio.bib}

\end{document}